\documentstyle[aps,preprint]{revtex}
\tightenlines

\begin{document}

\title{Local magnetic structure due to inhomogeneity of interaction in 
$S=1/2$ antiferromagnetic chain }

\author{Masamichi Nishino}

\address{Department of Chemistry, Graduate School of Science,\\
Osaka University, Toyonaka, Osaka 560, Japan\\} %

\author{Hiroaki Onishi and  Pascal Roos}
\address{Department of Earth and Space Science, Graduate School of Science,\\
Osaka University,       Toyonaka, Osaka 560, Japan}%

\author{Kizashi Yamaguchi}

\address{Department of Chemistry, Graduate School of Science,\\
Osaka University, Toyonaka, Osaka 560, Japan\\} %

\author{Seiji Miyashita}
\address{Department of Applied Physics, University of Tokyo\\
Bunkyoku, Tokyo, Japan}

\maketitle

\begin{abstract}
We study the magnetic properties of $S=1/2$ antiferromagnetic Heisenberg
chains with inhomogeneity of interaction.  
Using a quantum Monte Carlo method and an exact diagonalization method, 
we study bond-impurity effect in the uniform $S=1/2$ chain and also 
in the bond-alternating chain. Here {\lq}bond impurity{\rq} means a bond 
with strength 
different from those in the bulk or a defect in the alternating 
order. 
Local magnetic structures induced by bond impurities
are investigated both in the ground state and at finite temperatures, 
calculating the local magnetization, the local susceptibility and the local 
field susceptibility. 
We also investigate the force acting between bond impurities and 
find the force generally attractive.
\end{abstract}

\section{Introduction}

Quantum effects are very sensitive to the inhomogeneity of interaction. 
The magnetic structure induced by such inhomogeneity has much 
attracted us for its peculiar properties.
The effect of impurities for the $S=1/2$ Heisenberg antiferromagnetic 
chain 
has been studied extensively.
Eggert and Affleck~\cite{Affleck_PRB,Affleck_PRL} have studied the 
open chain,
where edges cause the inhomogeneity. 
They investigated the static structure function and the 
local susceptibility, using the conformal field theory and a Monte Carlo 
method. 
Laukamp, et al. have also studied the open chain problem 
by a Monte Carlo method and a DMRG method~\cite{Laukamp}.
Magnetic structures at edges cause interesting effects on the NMR line 
shape\cite{Takigawa,Uhrig}.
In particular, the observation\cite{Takigawa} of a broad background in 
the NMR spectrum in $\rm{Sr_2CuO_3}$
has been found to agree with the features predicted by the theoretical 
work on the field-induced local staggered magnetization~\cite{Affleck_PRL}.
The temperature dependence of the susceptibility of an ensemble of open 
finite chains with various lengths has been also studied experimentally 
and theoretically~\cite{Katsumata}.

The properties for $S=1/2$  chains with random exchange coupling have 
been also studied.
Random distribution of the interaction brings a new type of low 
temperature phase such as random singlet phases~\cite{Fisher} and 
furthermore various other types of randomness-induced phases have 
come out according to the 
distribution of the bonds~\cite{Furusaki_PRL,Furusaki_PRB}.

Impurity effect in the $S=1$ AF chain has been also investigated.
The existence of the edge state is one of the most interesting properties 
of the Haldane state~\cite{Haldane} which was exactly pointed out in the 
AKLT~\cite{AKLT}
chain. 
This edge state has been studied in detail by numerical
method~\cite{Kennedy,Miyashita}, where  the singlet and triplet 
(Kennedy triplet) states become
degenerate in the thermodynamic limit (the four-fold degeneracy).
Furthermore a doping of an impurity of $S=1/2$ brings a local magnetic 
structure which causes an interesting energy structures~\cite
{Hagiwara,Sorensen,Kaburagi}.
The effects of bond impurity has been also studied~\cite{Wang}.
There is the robustness of the local structure due to the $S=1/2$ impurity,
while the interaction between the local structures has been pointed 
out to be very weak because of the quasi degenerate energy 
structure~\cite{Pascal}.

In this paper we study magnetic structures in $S=1/2$ chains due to 
various spatial configurations of one or two bond impurities in the 
uniform system. 
Here we consider a uniform antiferromagnetic Heisenberg chain and 
we regard a weaker or stronger bond than the bulk bonds as an impurity 
bond.
First we review the properties of the uniform antiferromagnetic 
Heisenberg
model in open chains and investigate the temperature dependence
in detail.
When we put an impurity in an open uniform chain, 
it is found that such an impurity effectively separates the system 
and the right domain and the left domain behave 
almost independently at modestly low temperatures.
According to whether the number of spins of the domain is even or odd,
magnetic property of the domain shows very different characteristics.
We investigate the magnetic properties by calculating 
the magnetization profile, the spin correlation function,
the local susceptibility and the local field susceptibility (see next 
section).

We also study bond-impurity effects in the bond-alternating chain, where 
{\lq}impurity{\rq} means a defect of the alternating order. 
In the bond-alternating system, the correlation length is finite and 
similar to the case of $S=1$ Haldane systems.  
The impurity induces a localized magnetic structure around it.

In some cases the position of the bond impurity can move. 
We study the force acting between the impurity bonds. 
We find the force is attractive.

This paper is organized as follows.
In the next section, we explain briefly the method used in this study.
In Sect. \ref{sec:uni}, we investigate magnetic structures for 
the uniform systems with impurities.  
In Sect. \ref{alternate}, the effects of bond impurities for the magnetic 
structures of bond-alternating chains
are studied.  
In Sect. \ref{force}, we study the force between bond impurities.
Sect. \ref{sec:summary} is devoted to the summary and discussion.

\section{Model and Method}
In this paper we study the low temperature properties of 
$S=1/2$ antiferromagnetic Heisenberg chains with bond impurities.
The Hamiltonian is generally given as
\begin{equation}
{\cal H}=\sum_{i} J_i{\bf S}_{i}\cdot{\bf S}_{i+1}-\sum_{i}
h_i S_i^z, 
\end{equation}
where ${\bf S}_{i}=(S_i^x, S_i^y, S_i^z)$ are the $S=1/2$ spin operators.
In Sect III, we study the uniform chain where $J_i=J$ except at the position of the impurity bonds, $j$, where $J_j$ is different 
from the bulk value $J$. Hereafter we take $J$ as the unit of the energy.
In Sect IV, we study the bond-alternating chain where $J_i$ takes a 
strong bond $J_1$ and a weak bond $J_2$ alternately.
There the impurity means defects of the order of the alternation.

We mainly use the loop algorithm with continuous time quantum 
Monte Carlo method (LCQMC)~\cite{loop} in this study.
This method overcomes the problem of long autocorrelation in Monte 
Carlo update.
Furthermore replacing discrete time with finite Trotter number,
an algorithm using continuous time has been introduced~\cite{cont}
and the nuisance of the extrapolation of the Trotter number has been 
released.
These improvements allow us to study systems at very low
temperatures~\cite{onishi}.  In the present work, we performed 
$10^6$ Monte Carlo steps (MCS).  Here a MCS means a update of whole 
spins.

In the World Line quantum Monte Carlo method (WLQMC) we can control the value of total magnetization ($M_z$) in the initial state and
keep it by suppressing the global flip which changes the magnetization.
Thus, generally it is not difficult to obtain the true ground state
configurations fixing the value of the total magnetization.
However, in a standard LCQMC the number of world line and 
the winding number are updated automatically and we can not specify the value of
the magnetization $M_z$.
We adopt LCQMC with fixed total $M_z$ values.
We perform the standard LCQMC and store the data separately according to $M_z$.
When we need information for a specific value of $M_z$, we
use the data with that value of $M_z$ only.
This method worked successfully in the study of the site impurity
problem ($S=1/2$ spin) in the $S=1$ Heisenberg antiferromagnetic 
chain~\cite{Pascal}.
In order to check the method in the case of $S=1/2$ chains,
we confirmed the agreement between  the low temperature results 
obtained by the LCQMC for small sizes 
such as $L=20$ and 21 and those obtained by 
exact diagonalization. 

We study the local magnetic properties by investigating the following
properties: the local magnetization 
\begin{equation}
m_i=\langle S_i^z\rangle, 
\end{equation}
and the spin correlation function 
\begin{equation}
C(i, j)=\langle S_i^z S_j^z\rangle. 
\end{equation}

At finite temperatures, no local magnetization appears, $m_i=0$.
For the purpose of detecting the local magnetic structure, we calculate the local susceptibility introduced 
by Eggert and Affleck:

\begin{equation}
\left. \chi_{i} \equiv \frac{\partial}{\partial h} \langle S^z_{i}\rangle 
\right |_{h=0} = \beta \sum_{j} \langle S^z_{j} S^z_{i} \rangle
, 
\label{chi}
\end{equation}
where $h$ is the uniform field ($h_i=h$) and $\beta=1/T$.
This quantity is nonzero even at finite temperatures where all $M_z$
subspaces contribute. For the odd-chain (a chain with odd number of spins) $\chi_{i}$ is proportional to the local magnetization $m_i$ at $T=0$.

We also study another type of local susceptibility, which we call 
local field 
susceptibility in this paper.
\begin{equation}
\left. \chi_i^{local} \equiv \frac{\partial}{\partial h_{i}} \langle
S^z_{i}\rangle \right |_{h_{i}=0}=\int_{0}^{\beta} \langle S^z_{i}(\tau)
S^z_{i} 
\rangle d\tau -\beta {\langle S^z_{i} \rangle}^2, 
\label{tchi}
\end{equation}
which has been investigated as an indication of quantum fluctuation 
in the study of quantum spin glasses or diluted models~\cite{Ikegami}.

\section{bond impurities in uniform open chains}
\label{sec:uni}

In this section we study the effect of bond impurities in the uniform 
$S=1/2$ chain.
First, we review the magnetic properties of the open chains and 
investigate the temperature dependence of them. Next, we investigate magnetic properties in open odd-chains with one bond impurity.  And then, we investigate systems with two impurities and
how magnetic properties are changed by shifting the position of the two
impurities.

\subsection{Open uniform chain}
Eggert and Affleck~\cite{Affleck_PRB,Affleck_PRL} studied 
characteristics of magnetic properties for open uniform even-chains 
at low temperatures.
They found that local susceptibility remains near the edges for 
even-chains
at modestly low temperatures.
Laukamp, et al.~\cite{Laukamp} recently studied the magnetic properties for odd- or even-chains in a fixed $M_z$ space by DMRG. They found that 
the local 
susceptibility forms a simple sinusoidal form.
  These features are relevant to study the 
temperature dependence of experimental results. 

Fig. \ref{figchi_short_tlow}(a) and (b) show $\chi_{i}$ at a very low
temperature $T=0.01$ of chains with $L=63$ and 62, respectively.
The odd-chain shows a nodeless shape
of $\chi_i$ while $\chi_i$ of the even-chain shows a node, which reproduce
the result of Laukamp, et al. As the temperature increases,
$\chi_i$ at middle of the chain shrinks as shown in Fig. \ref{figchi63_005}
($L=63$, $T=0.05$).
In a longer chain, the shape reproduces
that of Eggert and Affleck\cite{Affleck_PRB,Affleck_PRL}
( $L=128$, $T=0.067$, Fig. \ref{figchi_long} (a)).
At this temperature $\chi_{i}$ of an odd-chain ($L=127$) shows a very
similar shape to that of $L=128$ as shown in Fig. \ref{figchi_long} (b).
On the other hand at the very low temperature $T=0.01$,
the profiles of $\{\chi_i\}$ for odd-chain and even-chain are very different
as shown in Fig. \ref{figchi_long} (c) and (d).
These profiles are essentially the same as those
in Figs. \ref{figchi_short_tlow}.

Here it should be noted that the amplitude of Fig. \ref{figchi_short_tlow}(b) is very small. 
In Fig. \ref{figchi62_t0.05} we show the data of $\chi_i$ at $T=0.05$ where
we find almost the same shape but the amplitude is about 20 times larger than 
that of $T=0.01$.
This temperature dependence is understood as follows.
In the subspace of the total magnetization $M_z(=\sum_{i} m_i)=0$, 
the local susceptibility vanishes by 
the definition Eq. (\ref{chi}). 
In the ground state we expect that $\chi_i$=0 assuming the ground 
state is singlet.
If a triplet state exists above it (with the energy gap $\Delta E$) and 
the other states have high energies and do not contribute to the 
thermal distribution, the amplitude of the shape is given by  
\begin{equation}
\chi_i=\gamma ~\overline{\chi_i}(M_z=1),
\label{amplitude}
\end{equation}
where 
\begin{equation}
\gamma = {2e^{-\beta\Delta E} \over 1+ 3e^{-\beta\Delta E}},
\label{gamma}
\end{equation}
and $\overline{\chi_i}(M_z=1)$ 
is the local susceptibility in the lowest level of $M_z=1$.
We  suppose that $\overline{\chi_i}$ is proportional to that in Fig. \ref{figchi_short_tlow}(b).

This ratio $\gamma$ is obtained by the fraction of MCS in the subspaces of $M_z=\pm 1$.
At $T=0.01$ the fraction is about 0.34 $\%$ and all other MCS are in the subspace of $M_z=0$.
Thus the above assumption that only the lowest singlet state and triplet 
states contribute to the thermal distribution is confirmed to be valid.

At $T=0.05$ the fraction of $M_z=\pm 1$ is 
observed to be about 34 $\%$ in MC simulation and small number of MCS distribute in the spaces of $|M_z|\ge 2$.
Thus the distribution is about 100 times larger than that at $T=0.01$,
which is consistent with the ratio, $\chi_i(T=0.05)/\chi_i(T=0.01)
\simeq 20,$ taking into account 
$\beta$ in the definition of $\chi_i$ (Eq. (\ref{chi})).

Using the relation Eq. (\ref{gamma}), from this fraction $\gamma=
0.0034$ at $T=0.01$ the energy gap 
is estimated as 
\begin{equation}
\Delta E(L=62) \simeq 0.064.
\end{equation}
In the chain of $L=128$ the energy gap is similarly estimated as
\begin{equation}
\Delta E(L=128) \simeq 0.031.
\end{equation}
The energy gap becomes small as the size increases proportionally to 
$1/L$ as is expected for $S=1/2$ antiferromagnetic Heisenberg chains.

In Fig. \ref{figchi2_62_63}(a) and (b), $\chi_i^{local}$ for $L=63$ and 62
are shown, respectively. In both cases, $\chi_i^{local}$ also shows a zigzag shape, although in Fig. \ref{figchi2_62_63}(b) there is a node 
as in Fig. \ref{figchi_short_tlow} (b). 
This zigzag behavior indicates that the amplitude of quantum fluctuations
is larger at sites where the local spin is antiparallel to the field.

\subsection{One bond impurity}

As bond impurities we consider three types of inhomogeneity:
(a) a strong bond impurity ($J_{\rm s}$=2),
(b) a weak bond impurity ($J_{\rm w}$=0.5), and
(c) a complex structure of impurities consisting of both strong
($J_{\rm s}$=2) and week ($J_{\rm w}$=0.5) bond impurities.
Exactly speaking, the last type does not consist of a single bond impurity but we investigate this case here because of an analog to the formers.
In Fig. \ref{figmag_oneimpu}(a), (b) and (c), 
the magnetic profiles $\{m_i\}$  obtained by LCQMC at $T=0.01$ in the $M_z=1/2$ space
are shown for the types. Configurations of bonds are also illustrated in the figures.
These profiles are considered to represent well the ground state 
because the structure shows little dependence on the temperature within 
a fixed value of magnetization ($M_z$).  
In Fig. \ref{figmag_oneimpu}(a) the left hand side of the impurity shows 
a steady staggered structure, 
while there remains little magnetization in the right hand side. 
This structure is easily understood from the viewpoint of
the Lieb-Mattis theorem~\cite{Liep}.
$S=1/2$ open odd-chain has the doublet $M_z=\pm 1/2$ state as the ground state and the local magnetization is nonzero, while
the even-chain has the singlet $M_z=0$ ground state and 
no local magnetization appears.
In the configuration of Fig. \ref{figmag_oneimpu}(a), 
the spins connected by the strong bond impurity form a singlet 
state and the whole system is efficiently divided into two parts consisting
of spins 1st-29th (odd) and 32nd-61st (even). 
Thus, approximately, the doublet ground state appears in the left 
domain and
the singlet ground state appears in the right domain.  
On the other hand in the configuration of Fig. \ref{figmag_oneimpu}(b), 
the weak bond itself cuts the whole chain and the system is 
divided into two part: 1st-30th (even) and 31st-61st (odd).
Consequently the structures of the right hand side and the left hand side 
exchange.
Due to the same reason, a magnetic structure also appears in (c), where 
the two domains are more clearly separated and the contrast between 
the right and left sides is more clear.

In order to investigate the correlations in each domain, we calculate the
two-point correlation functions of the model of Fig. \ref{figmag_oneimpu} (a) in the fixed $M_z=1/2$ space at $T=0.01$. 
The two-point correlation functions from the left edge $C(1, i)$ and the right edge $C(i, L)$ 
are compared in Fig. \ref{fig_oneimpu_corr}(a) and the two-point correlation functions 
from the
middle site of the left domain $C(15, i)$ and the right domain $C(i, 47)$ are compared in Fig. \ref{fig_oneimpu_corr}(b). 
While the magnetic profile $\{m_i\}$  of each domain in the ground state is very 
different from each other, almost the same correlation in both domains 
exists which is also very similar to that in the pure $S=1/2$ chain.
Because the correlations are almost the same, we expect that the spins in the right domain fluctuate coherently keeping the mutual correlation.
That is to say, the spins in the right domain have a similar magnetic 
profile to that in the left domain in some direction and the direction is fluctuating.
The same tendency is also observed in the models of Fig. \ref{figmag_oneimpu} (b) and (c).

We also investigate 
$\chi_{i}$ and $\chi_{i}^{local}$ of the lattice with one strong bond 
impurity for $L=127$, where the impurity bond locates between 
64th and 65th sites.
Here the left and right domains contain 63 and 62 sites, respectively.
When the temperature is modestly low such as $T=0.05$ we find that each domain independently shows magnetic property according to whether 
the number of spins is odd or even, as shown in Fig. 
\ref{figchi_one_0.05}.
At a very low temperature such as $T=0.01$,
$\{\chi_{i}\}$ in the right hand side is expected to be given by that 
in Fig \ref{figchi_short_tlow} (b) where the amplitude is very small with 
a node, 
if the domain behaves independently.
On the contrary, the amplitude of $\chi_i$ in the right hand side is enhanced and the one-node structure disappears,
as shown in Fig.  \ref{figchi_one_0.01}. This change is understood as follows.
If the domain would behave completely independently, $\chi_i$ is 
nearly zero because the singlet ground state dominates 
and the amplitude of the one-node structure decreases
as Eq. (\ref{amplitude}) as we see in the previous subsection.
However in the present lattice, the right domain can be regarded as 
an even-chain only approximately because of nonzero interaction through the impurity bond. 
Thus the magnetization of the domain is no more a good quantum number. 
Consequently in the lowest state of the right domain, $\chi_i$ may have some structure.
The structure is expected to have small local magnetic moment 
because the state would be nearly singlet.
In Fig. \ref{figchi_one_0.02}, we show $\chi_i$ at an intermediate temperature.
Although we do not show it here, the magnetic profile $\{m_i\}$ is 
found to be proportional to $\{\chi_i\}$ in this lattice.

The local field susceptibility at $T=0.05$ is shown in Fig. \ref{figchi_one_0.05} (b).
When we decrease the temperature to $T=0.01$, $\{\chi^{local}_i\}$ is found 
to change smoothly to a shape 
given by the combination of Fig. \ref{figchi2_62_63} (a) and (b) in 
each domain.

\subsection{two bond impurities}

In this subsection we consider two-impurity problem.
There are various configurations with two bond impurities 
concerning to combinations of types of impurities and also 
to whether the number of spins in each domain is odd or even.  
Here we deal with two types of lattices of $L=61$ ($J=1$) with two strong bond impurities ($J_{\rm s}=2$).
The impurities divide the lattice into three domains:    
model (A) a lattice where the domains contain 
successively odd, even, and even number of spins  
and model (B) a lattice where the domains contain odd, odd, and odd number of spins.
The lattice structures and the magnetic profiles at $T=0.01$ are presented 
in Fig. \ref{fig_twoimpu_mag}(a) and (b).
These magnetization profiles represent well the ground states of the models 
because almost all (100 and 98.9$\%$) of Monte Carlo steps are distributed 
in the $M_z=\pm1/2$ space which means the system is in a doublet state.
The feature of these magnetic profiles can be understood from the analogy
to the arguments given in the previous section.
Approximately, the doublet ground state appears in domains with 
odd number of spins, while the singlet ground state appears in domains with even number of spins.

For the model  (B), 
in a naive picture we may consider the magnetizations of domains to 
be +1/2, $-$1/2 and +1/2. Hereafter we denote this configuration by (1/2,$-$1/2,1/2).
Fig. \ref{fig_twoimpu_mag_sum} shows the summation of magnetization per site from the left edge site 
\begin{equation}
{\cal M}_z(j)=\sum_{i=1}^j m_i, 
\end{equation}
which shows that the right and left domains have positive
magnetizations, while the middle domain 
has a negative magnetization.
In this figure we find tendency of a staggered domain 
magnetization. 
The magnetizations of the domains are reduced to 0.3, $-$0.1, and 0.3, respectively.
This reduction comes from a linear combination of the state (1/2, 1/2, 
$-$1/2)
and ($-$1/2, 1/2, 1/2) due to the quantum fluctuation.
Looking on the domain magnetization as an effective $S=1/2$ spin
interacting by an exchange $\tilde{J}$, 
this system is modeled by a 
three-site Heisenberg model ${\cal H}=\tilde{J}{\bf S}_{1}\cdot{\bf S}_{2}+
\tilde{J}{\bf S}_{2}\cdot{\bf S}_{3}$. 
In the $M_z=1/2$ space the eigenvalues and the eigenvectors are described as 
\begin{eqnarray}
&& E_{1}=-\tilde{J},  \nonumber \\
&& |\phi_{1}\rangle =1/\sqrt{6}(|++-\rangle -2 |+-+\rangle +|++-\rangle) \nonumber \\
&& E_{2}=0,  \nonumber \\
&& |\phi_{2}\rangle =1/\sqrt{2}(|++-\rangle -|-++\rangle) 
\label{ground eq} \\
&& E_{3}=\tilde{J}/2,  \nonumber  \\
&& |\phi_{3}\rangle =1/\sqrt{3}(|++-\rangle + |+-+\rangle +|-++\rangle). \nonumber 
\end{eqnarray}
Here, we find $\langle \phi_{1}| S_{1}^z| \phi_{1}\rangle=1/3$ and 
$\langle \phi_{1}| S_{2}^z| \phi_{1}\rangle=-1/6$. The reduced magnetization of the domains is close to this set 
(1/3, $-$1/6, 1/3) and thus the domain structure is well described by the state $\phi_1$.  
We find that the state of each domain can be regarded as a doublet state approximately.

Finally in Fig. \ref{chi_two005} we show $\chi_i$ at modestly low temperature ($T=0.05$) for a long chain of the model (A), where we find again the independent behavior of
the domains.

\section{bond impurity in bond-alternating chains}
\label{alternate}

In this section, we investigate models with bond alternation, 
$\cdots J_{1} J_{2} J_{1} J_{2}   \cdots$, where $J_{1} > J_{2}$. 
Here we study the effect of a defect of the alternation, such as 
$\cdots  J_{1} J_{2} J_{1} J_{2}  \underline{J_{1} J_{1}} J_{2} J_{1} J_{2}
J_{1}  \cdots $.
 As mentioned in the introduction, the alternating chain 
has a energy gap between the ground state and the first excited state and 
the spin correlation length is finite.  
These features are similar 
to those in the $S=1$ AF Heisenberg model which is the Haldane system.  
As well as in the case of the Haldane system, there are edge 
states in this model, too.  The edge state strongly depends on the bond 
situation at the edge.  Thus we study the following four cases of the
configurations of the bonds at the edges and the center:\\
(a) the two strong bonds are at the center and the edges terminate 
with strong bonds,\\
(b) the two strong bonds are at the center and the edges terminate 
with weak bonds,\\
(c) the two weak bonds are at the center and the edges terminate 
with weak bonds, \\
and \\
 (d) the two weak bonds are at the center and the edges terminate 
with strong bonds.\\
Here we take the strong bond to be $J_{1}=1.3$ and the weak bond to be
$J_{2}=0.7$. 
The magnetic profiles of (a)-(d) are drawn in Figs. \ref{fig_bondalt_mag}, 
where simulations were performed at $T=0.01$ and in the $M_z=1/2$ space. 
A magnetization is induced locally around the impurity. 

First we consider the cases where the bonds at edges are strong, i.e., 
model (a) and (d).
If we allocate a singlet pair at each strong bond, a structure with
 neighboring two strong bonds remains at the center in the model (a), while one site remains in the model (d). The magnetization of $M_z=1/2$ is assigned in the remaining part of the lattices to induce a local magnetic structure.
Because the edge bonds are strong, no magnetization is induced at the edges.  The ground state is simple doublet and 100$\%$ of Monte Carlo steps are distributed in the $M_z=\pm1/2$ space.
Fig. \ref{fig_bondalt_mag}(a) and (d) are considered to describe well the 
magnetization profiles of the ground state 
of both models.  
 In the case 
(a) negative magnetization appears at the middle site, while
in the case (d) positive  magnetization appears there.
This structure is naturally understood as follows.
The interaction of the three spins at the center of the model (a) is represented as three-spin model coupled by the strong bonds where 
the state $ |\phi_{1}\rangle$ in Eq. (\ref{ground eq}) gives the ground state, while in the model (d) a spin at the center is isolated from 
the others which form singlet states.

On the other hand, in the model (b) and (c), 
if we allocate singlet spins at the strong bond, spins remain 
at the center and  also at both edges and thus
there are three positions for 
magnetic moments. 
Indeed in  both (b) and (c) models, magnetization is
induced locally around the impurity and the right and the left edge sites. 
Fig. \ref{fig_bondalt_sum} shows the summation of magnetization per site from the left edge
site for the model (b). In Fig. \ref{fig_bondalt_sum} the values of the left plateau and right plateau are 0.166 and 0.333, respectively. 
From this figure we find a spin 1/6 locates at each local structure. This deceptive fractional magnetization is considered to come
from mixing of states.
As we mentioned in Sect. III, the eigenstates in $M_z=1/2$ of the open chain of three 
spins are given by Eq. (\ref{ground eq}), where

\begin{eqnarray}
&& \langle \phi_{1}|S_{1}^z|\phi_{1}\rangle=\langle \phi_{1}|
S_{3}^z| \phi_{1}\rangle=1/3 \ {\rm and} \nonumber \\
&& \langle \phi_{1}|S_{2}^z|\phi_{1}\rangle=-1/6, \nonumber \\
&& \langle \phi_{2}|S_{1}^z| \phi_{2}\rangle=\langle \phi_{2}|S_{3}^z| 
\phi_{2}\rangle=0  \ {\rm and} \\
&& \langle \phi_{2}|S_{2}^z|\phi_{2}\rangle=1/2,  \nonumber \\
&& \langle \phi_{3}|S_{1}^z| \phi_{3}\rangle=\langle \phi_{3}|
S_{2}^z| \phi_{3}\rangle=\langle \phi_{3}|S_{3}^z| \phi_{3}\rangle=1/
6. \nonumber  
\end{eqnarray}
Although there are energy gaps between the state $|\phi_{1}\rangle$, 
$|\phi_{2}\rangle$, and $|\phi_{3}\rangle$, the temperature is considered to be much larger than these energy gaps and these states are almost 
degenerate and thus the expectation values of $S_{1}^z$ and $S_{2}^z$ are 
(1/3+0+1/6)/3=1/6 and ($-$1/6+1/2+1/6)/3=1/6, respectively. 

In order to check that the ground state of the type (b) is represented 
by the $| \phi_{1}\rangle$ of the three-spin model, 
we investigate a short chain of $L=21$ by the diagonalization method. 
Because the length of the chain is short, we choose a shorter 
localization length. For this purpose, we set strong and weak bonds $J_{1}=2$ and $J_{2}=0.5$,
respectively.
Fig. \ref{fig_bondalt_diagmag} (a) shows the magnetic structure in the ground 
state of this model
and Fig. \ref{fig_bondalt_diagmag} (b) shows the summation of the magnetization per site from
the left edge site.
The net magnetization around the right and the left edge is
positive and that around the impurity site is negative.  
The value for the left plateau of the Fig. \ref{fig_bondalt_diagmag}(b) is about 1/3 and this
corresponds to $\langle \phi_{1}|S_{1}^z| \phi_{1}\rangle$, and the 
value of the second plateau is about 1/6, which also corresponds to 
$\langle \phi_{1}|S_{1}^z| \phi_{1}\rangle$ + $\langle \phi_{1}|S_{2}^z|\phi_{1}\rangle$.
Thus the three-spin model is valid for this model of $L=21$. 
On the other hand in the case of $L=63$, the energy gap is much smaller 
than the temperature. Thus all the quasi-degenerate states contribute 
to the observation. 

This degeneracy is also seen in the distribution of $M_{z}$.
$M_{z}$ distributes as shown in Fig. \ref{fig_bondalt_prob}. Here the distribution for $M_z=3/2$ is about 12.5 $\%$.  
There are three states in the $M_z=1/2$ space and one state in the $M_z=3/2$ space in the three-spin model.
The fact that the distribution for total $M_z=3/2$ is about 12.5 $\%$ means that
$T=0.01$ is much higher compared with the energy gaps between these four states.  
In the three-spin model the eigenvalue of the state of $M_z=3/2$ is $\tilde{J}/2$, while the eigenvalues of the state of $M_z=1/2$ are $-\tilde{J}$, 0, and $\tilde{J}/2$ (Eq. (\ref{ground eq})).
In principle we can obtain the energy gap from the temperature 
dependence of the distribution.  However it is too small to be detected here.
Thus these three states in the $M_z=1/2$ space appear in equal
probability, and $\langle S_{1}^z \rangle =\langle S_{2}^z \rangle =\langle S_{3}^z \rangle =1/6$ in the $M_{z}=1/2$ space as mentioned above.
In the $M_z=3/2$ space we observe  $S=1/2$ moment at each local structure as
shown in Fig. \ref{fig_mag_Mz3/2}(a) and (b).
We find a similar scenario for the model (c).
Thus we conclude that in rather strong bond-alternating systems, 
local magnetic structures induced by a bond impurity or weak edge 
bonds have an effective $S=1/2$ spin and they behave almost
independently.

\section{Force between bond impurities}
\label{force}
If we allow the position of impurities to move, impurities move by the 
force between them.
In the equilibrium state, the distance between impurities distributes in 
the canonical distribution 
for the interaction energy between the impurities.
Here we study the force between the bond impurities.

\subsection{Two bond impurities in the uniform chain}
In this section we consider the force between bond 
impurities in the uniform chain.
First, we deal with $L=12$ periodic chains 
($J=1$) including two strong 
bond impurities ($J_{\rm s}=2$) at various positions.
We study the ground state energy of the system as a function 
of the position of impurities.
An impurity is put on the first bond and the other is put on the $n$-th bond.
We calculate energies of the system      
as a function of the distance 
\begin{equation}
\Delta=n-1,
\end{equation}
 which is shown in Fig. \ref{fig_force_diag}.
The energy $E(\Delta)$ corresponds to the potential energy of the 
force between the impurity bonds.
We find that the cases in which $\Delta$ is even are more 
energetically favorable than
those of odd $\Delta$.
Among the cases of even-distance the energy is lowest when the 
distance is shortest, which indicates 
that an attractive force acts between the bond impurities. 

To confirm this conclusion for the attractive force in a larger system, 
we preformed Monte Carlo simulations for an $L=60$ periodic chain.
Here we use the following algorithm to update the system.
We start with an arbitrary configuration with two impurities, and then
we update the spin variable with the standard LCQMC.
Next we update the bond configuration in the following way. 
We exchange the 
neighboring bonds sequentially from the left.
If the two bonds are the same, we skip to the next. We choose a new 
configuration of bonds by the thermal 
bath algorithm, i.e., we take the exchanged configuration with the 
probability:
\begin{equation}
p=\frac{W'}{W+W'}
\end{equation}
where $W$ is the Boltzmann weight for the original bond configuration 
and $W'$ is that of the exchanged configuration for the given spin 
configuration $| \sigma \rangle $:
$W$=$\langle \sigma | e^{-\beta \cal{H_{\rm B}}} | \sigma
\rangle $ and  
$W'$=$\langle \sigma | e^{-\beta \cal{H_{\rm B'}}} | \sigma 
\rangle $.
We plot the distribution of the distance between 
the impurity bonds in Fig. \ref{figforce60}.  
We expect that the distribution should be the canonical distribution 
with the potential energy of the interaction of 
the impurities. We 
see in Fig. \ref{figforce60} that the most probable distance is $\Delta=2$. Thus we conclude that $\Delta=2$ gives the lowest energy. 
When $\Delta$ is odd, 
the probability is low, which indicates 
that the energies for odd-distances are higher than those for even-distances. 
Among the even $\Delta$s, the probability decreases as the 
distance becomes long, which indicates the attractive force acts.

We also studied the case of weak impurity bond and found that the attractive force acts similarly to the case of the strong impurity bonds.

\subsection{Two defects in an alternating chain}
In the alternate chain ($ \cdots J_{1} J_{2} J_{1} J_{2}  \cdots$ ), 
the system may have a pair of defects by shifting a position 
of a strong  bond by one. 
\begin{equation}
\cdots J_{1} J_{2} \underline{J_{1} J_{1} J_{2} J_{2} } J_{1} J_{2} 
\cdots .
\label{ichi}
\end{equation}
If we shift the position furthermore, the system has a configuration 
\begin{equation}
\cdots \underline{J_{1} J_{1}} J_{2} J_{1} \underline{ J_{2} J_{2} } 
\cdots , 
\label{ni}
\end{equation}
etc. We study dependence of the energy on the distance ($\Delta$) between the positions of $J_{1} J_{1}$ and $J_{2} J_{2}$. In the case of no defect we
define $\Delta= 0$, $\Delta=1$ for Eq. (\ref{ichi}),  $\Delta=2$ for Eq. (\ref{ni}) and so on.  In Fig. \ref{fig_force_alt_diag} 
we plot the ground state energy as a function of $\Delta$
obtained by exact diagonalization for $L=24$ with $J_1=2$ and 
$J_2=1$.
Here we find again an attractive force between the impurities.

\section{Summary and Discussion}
\label{sec:summary}

In this paper we explored effects of various bond impurities on the
low temperature magnetic properties in the 
uniform and bond-alternating antiferromagnetic Heisenberg chains. 

In particular we clarified the temperature
 dependence of the magnetic
structure of the open uniform chain. 
At modestly low temperatures systems show a similar local 
susceptibility profile regardless of whether the number of spins 
is even or odd. On the other hand at very low temperatures, 
the profile is very different from each other.

We also studied domain structures which are 
separated by bond impurities in the uniform chain. There we found that each domain behaves almost
independently at modestly low temperatures. 
However, at very low temperatures in the domain with even number of spins, a magnetic structure appears due to the 
interaction between the domains, although such structure does not exist in a completely isolated open chain with even number of spins.
Thus the interaction between the domains is not negligible at very 
low temperatures. 
We also found that the interactions between domains with odd number 
of spins can be approximately modeled by a simple system with $S=1/2$ spins representing the domain magnetization.
The effective interaction $\tilde{J}$ of the model is not necessarily 
small at $T=0.01$. Because the energy gap due to the finite size 
in the uniform chain is proportional to $1/L$, we expect that 
$\tilde{J}$ also becomes small as $1/L$.

Effect of bond impurities are investigated also in the bond-alternating chain.
The role of the bond impurity is very 
different from that in the uniform chain, i.e. bond impurities induce a magnetic
structure around them, while they cause the division into domains in the 
uniform chain.
In the bond-alternating chain the effect is quite similar to that of the site impurities of $S=1/2$
in the $S=1$ antiferromagnetic chain because of the gapful nature.
Here induction of magnetic moments at edges depends on whether the edge bonds are strong or weak.   
In the bond-alternating chain, the magnetic structures behave 
almost independently even at very low temperatures, 
contrary to the case of the uniform chain.
If we model the interaction of the magnetic structure 
by a effective spin model, the interaction $\tilde{J}$ is expected to be 
exponentially small with the system size, which is due to 
the gapful nature of the bond-alternating chain.

We also investigated the force
acting between the bond impurities and studied the distribution of 
impurities when impurities are allowed to move.  
It turned out that the force is attractive both between the impurity 
bonds in the uniform chain, and between the defects of alternation 
in the bond-alternating chain.

We hope that the present study helps to analyze the magnetic properties 
at low temperatures such as observed by NMR measurement.

\acknowledgements

The present authors would like to thank Professor Jean-Paul Boucher
for his valuable and encouraging discussion.
The present work was supported by Grant-in-Aid for Scientific
Research on Priority Areas (No. 10149105 {\lq \lq}Metal-assembled Complexes{\rq \rq}
from Ministry of Education, Science, Sports and Culture of Japan.
M. N. was also supported by the Research Fellowships of the Japan 
Society for the Promotion of Science for Young Scientists.

\begin{figure}
\caption{Local susceptibility $\chi_i$ at $T=0.01$.
(a) $L=63$ and (b) $L=62$.}
\label{figchi_short_tlow}
\end{figure}

\begin{figure}
\caption{Local susceptibility $\chi_{i}$ at $T=0.05$ for $L=63$.
}
\label{figchi63_005}
\end{figure}

\begin{figure}
\caption{ Local susceptibility $\chi_i$ at $T=0.067$ 
(a) $L=128$ (b) $L=127$ and at $T=0.01$ (c) $L=127$ and (d) $L=128$.   
}
\label{figchi_long}
\end{figure}

\begin{figure}
\caption{Local susceptibility $\chi_i$ at $T=0.05$ for $L=62$.
}
\label{figchi62_t0.05}
\end{figure}

\begin{figure}
\caption{Local field susceptibility $\chi_i^{local}$ at $T=0.01$ for
(a) $L=63$ and (b) $L=62$.
}
\label{figchi2_62_63}
\end{figure}

\begin{figure}
\caption{Local magnetization in the $M_{z}=1/2$ space at $T=0.01$
for the models with $L=63$ (a)-(c): (a) strong bond impurity ($J_{\rm s}=
2$) 
(b) weak bond impurity ($J_{\rm w}=0.5$)
(c) strong ($J_{\rm s}=2$) and weak ($J_{\rm w}=0.5$) bond impurities.
}
\label{figmag_oneimpu}
\end{figure}

\begin{figure}
\caption{(a) Two-point correlation functions from the left edge and
the right edge are compared.
(b) Two-point correlation functions from the middle site of the left domain and the right domain are compared.
}
\label{fig_oneimpu_corr}
\end{figure}

\begin{figure}
\caption{(a) Local susceptibility of the lattice with one strong bond
impurity for $L=127$ at $T=0.05$, where the impurity bond locates 
between 64th and 
65th site. 
(b) Local field susceptibility of the lattice with one strong bond
impurity for $L=127$ at $T=0.05$. 
}
\label{figchi_one_0.05}
\end{figure}

\begin{figure}
\caption{Local susceptibility of the lattice with one strong bond
impurity for $L=127$ at $T=0.01$.
}
\label{figchi_one_0.01}
\end{figure}

\begin{figure}
\caption{Local susceptibility of the lattice with one strong bond
impurity for $L=127$ at $T=0.02$.
}
\label{figchi_one_0.02}
\end{figure}

\begin{figure}
\caption{(a) Lattice structure and local magnetization for the model (A).
(b) Lattice structure and Local magnetization for the model (B).
}
\label{fig_twoimpu_mag}
\end{figure}
 
\begin{figure}
\caption{
Summation of magnetization per site from the left
edge site for the model (B).
}
\label{fig_twoimpu_mag_sum}
\end{figure}

\begin{figure}
\caption{
$\chi_i$ at $T=0.05$ for a long chain ($L=127$) of the type of the model (A).
}
\label{chi_two005}
\end{figure}

\begin{figure}
\caption{
Local magnetization of the models (a)-(d) at $T=0.01$ in the $M_{z}=1/2$ space.
(a) and (c) contain 63 sites and (b) and (d) contain 65 sites.
The diamonds denote the strength of bonds $J_i$, those at high positions 
denote $J_1$ and those at the low positions denote $J_2$.
Details are shown in text.
}
\label{fig_bondalt_mag}
\end{figure}

\begin{figure}
\caption{
Summation of magnetization per site from the
left edge site for the model (b).
}
\label{fig_bondalt_sum}
\end{figure}

\begin{figure}
\caption{(a) Local magnetization of a bond-alternating chain with $L=21$, which is the same type as that in Fig. \ref{fig_bondalt_mag}. (b).
The strong and weak bonds are $J_{1}=2$ and $J_{2}=0.5$, respectively.
(b) Summation of the total magnetization per site from the left edge.
}
\label{fig_bondalt_diagmag}
\end{figure}

\begin{figure}
\caption{
Distribution of $M_{z}$  for the model (b) in Fig. \ref{fig_bondalt_mag}.
}
\label{fig_bondalt_prob}
\end{figure}

\begin{figure}
\caption{
(a) Local magnetization for the model (b) in Fig. \ref{fig_bondalt_mag}
in the $M_z=3/2$ space at $T=0.01$.
(b) Summation of the total magnetization per site from the left edge.
}
\label{fig_mag_Mz3/2}
\end{figure}

\begin{figure}
\caption{
Ground state energy as a function of
$\Delta$ obtained by exact diagonalization for $L=12$ periodic chains 
with two strong bond impurities.
}
\label{fig_force_diag}
\end{figure}

\begin{figure}
\caption{
Distribution of $\Delta$ by a Monte Carlo method 
for $L=60$. Open squares and circles show data for $T=0.1$ and $T=0.05$, 
respectively.
}
\label{figforce60}
\end{figure}

\begin{figure}
\caption{Ground state energy as a function
of
$\Delta$
obtained by exact diagonalization in $L=24$ for bond-alternating systems with
defects.
}
\label{fig_force_alt_diag}
\end{figure}

\end{document}